\documentclass[12pt]{article}
\usepackage[a4paper, total={6.1 in, 8.8 in}]{geometry}
\usepackage {helvet}
\usepackage{amsthm, amsmath, amssymb}
\usepackage{mathtools}
\usepackage{graphicx,subcaption,float}
\usepackage{breqn}
\usepackage[autostyle]{csquotes}
\usepackage[british]{babel}
\usepackage{breakcites}

\DeclarePairedDelimiterX{\norm}[1]{\lVert}{\rVert}{#1}
\title{A TEST FOR DETECTING STRUCTURAL BREAKDOWNS IN MARKETS USING EIGENVALUE DISTRIBUTIONS}

\date{\vspace{-5ex}}

\author{Malay Bhattacharyya\footnote{Address correspondence to Malay Bhattacharyya, Professor, Decision Sciences, IIM Bangalore, India; e-mail:malayb@iimb.ac.in} $^{1}$  and 	Rajesh Kasa$^{1,2}$ \\
	 $^{1}$Indian Institute of Management, Bangalore  \\
	 $^{2}$School of Computing, NUS, Singapore
		}

\begin{document}

\maketitle

\begin{abstract}
Correlations among stock returns during volatile markets differ substantially compared to those from quieter markets. During times of financial crisis, it has been observed that \textquote{traditional} dependency in global markets breaks down. However, such an upheaval in dependency structure happens over a span of several months, with the breakdown coinciding with a major bankruptcy or sovereign default. Even though risk managers generally agree that identifying these periods of breakdown is important, there are few statistical methods to test for significant breakdowns. The purpose of this paper is to propose a simple test to detect such structural changes in global markets. This test relies on the assumption that asset price follows a Geometric Brownian Motion. We test for a breakdown in correlation structure using eigenvalue decomposition. We derive the asymptotic distribution under the null hypothesis and apply the test to stock returns. We compute the power of our test and compare it with the power of other known tests. Our test is able to accurately identify the times of structural breakdown in real-world stock returns.  Overall we argue, despite the parsimony and simplicity in the assumption of Geometric Brownian Motion, our test can perform well to identify the breakdown in dependency of global markets.
\end{abstract}


\section{Introduction}
An important problem in statistical modeling of financial time series is to analyze and detect structural changes in the relationship among stock returns. Pearson Correlation is one of the widely used metrics in financial risk management to indicate the relationship among various returns. Long-term risk-averse investors tend to hold portfolios of assets whose returns are not positively correlated for diversification benefits. However, there is compelling empirical evidence that the correlation structure among returns of the assets cannot be assumed to be constant over time, see, e.g. \cite{forbes2002no}, \cite{krishnan2009correlation}, \cite{wied2012testing} and \cite{wied2017nonparametric}. In particular, in periods of  financial crisis, correlations among stock returns increase, a phenomenon which is sometimes referred to as diversification meltdown.  In this paper, we detect and test for these structural changes by considering the constancy of correlation matrix. \cite{wied2012testing} has shown that testing for changes of correlation can be more powerful than testing for changes in covariance, especially when there is more than one change point. However, one of the drawbacks of these existing tests (\cite{wied2012testing} and \cite{wied2017nonparametric})  is that pairwise comparison of correlation matrix is not a scalable solution when there are large number of stocks involved in the portfolio. Instead of the vector of successively calculated pairwise correlation coefficients, we consider the largest eigenvalue of sample correlation matrix and derive its limiting distribution, based on the assumption of Geometric Brownian motion (GBM) of stockprice  and some proof ideas from \cite{anderson1963asymptotic}. Our key contributions are as follows: Our proposed test is scalable to portfolios with large number of stocks as ours does not involve any pair-wise comparisons. It is easily explainable to all the stakeholders and fairly straightforward to implement. 

\paragraph{Outline}
First, we show how the returns are normality using the assumption of GBM for stock prices.  Next, we define our test statistic and derive its distribution using results from \cite{anderson1963asymptotic}. Our test statistic is qualitatively very similar to the one defined in \cite{wied2012testing} and \cite{wied2017nonparametric} as discussed in section \ref{sec:twostock}. Towards the end of this section, we give a simplified expression for the asymptotic distribution of test statistic for the cases of two-stock and three-stock portfolios. In section \ref{Testing on stock returns}, we demonstrate the performance of our test on known stock indices such as SNP500, DOWJones, etc. In section \ref{Local power}, we derive the power of our test and compare it with those of existing methods. Finally, in section \ref{conclusions}, we discuss the merits and demerits of our approach.

\section{Test statistic} \label{documentclasses}

In this section, we derive the distribution of eigenvalues of sample correlation matrix when the returns are iid multivariate normal by making use of results from \cite{anderson1963asymptotic}. Further, we define our test statistic and use these results to derive its asymptotic distribution. First, we start off by showing how the returns will be iid normal under the assumption that the stock price follows a Geometric Brownian Motion. 

\subsection{The Geometric Brownian Motion - preliminaries} 
The Geometric Brownian Motion is a continuous time stochastic process in which the logarithm of a random variable follows a Wiener's process with some drift \cite{sheldonross}. Geometric Brownian motion (GBM) describes the evolution of the stock price $S(t)$ over time and is widely used in mathematical finance to calculate the price of options \cite{gbmstochastic}. The GBM is modeled using the following stochastic differential equation:

\begin{equation}\label{eq:gbm1}
dS(t) =\mu S(t) dt +\sigma S(t) dW(t)
\end{equation}

Here, $\mu$ and $\sigma$ denote the drift and volatility of the stock price, respectively.  $W$ is a standard Brownian motion, also known as Wiener's Process, that is characterised by independent identically distributed (\textit{iid}) increments of random variables that follow a Gaussian distribution with zero mean and a standard deviation equal to the square root of the time step. It can be seen that at every time $t$, $dS(t)$ depends only $S(t)$. In other words, the conditional distribution of the future price given all the price information up to time $t$, depends only on the present price at time $t$ but not on the past prices - i.e. a Markov property.  

 Using It\^{o}s lemma, equation \ref{eq:gbm1} can be rewritten as follows:

\begin{equation}\label{eq:logS}
d \log S(t)=\left(\mu-\frac{1}{2}\sigma^{2}\right) dt +\sigma dW(t)
\end{equation}

where log denotes the standard natural logarithm. 
Integrating this equation (\ref{eq:logS}) from $t_1$ to $t_2$, leads to:

\begin{equation}\label{eq:logSdistrib}
\begin{split}
    \log S(t_2) - \log S(t_1)=\left(\mu-\frac{1}{2}\sigma^{2}\right) (t_2-t_1) +\sigma (W(t_2)- W(t_1)) \\ \sim N\left(\left(\mu-\frac{1}{2}\sigma^{2}\right) (t_2-t_1),\sigma^2(t_2-t_1) \right).
\end{split}
\end{equation}

Rearranging the equations and substituting the boundary conditions $t_2=T$ and $t_1=0$, the process describing the stock price $S(T)$ is obtained as follows:

\begin{equation} \label{logreturns}
S(T)=S(0)\exp\left(\left[\mu-\frac{1}{2}\sigma^{2} \right]T+\sigma W(T) \right)
\end{equation}

Equation (\ref{logreturns}) shows that if asset price $S(t)$ follows a GBM, then the logarithmic returns $\log(S_{t+\Delta t}/S_{t})$ follow a normal distribution.

Note that in equation (\ref{eq:logSdistrib}), we can approximate the log returns using Taylor series approximation of $\log(1-x) \approx - x + o(x)$

\begin{equation}\label{eq:taylorapprox}
    \log\left(\frac{S(t_2)}{S(t_1)} \right) = \log\left(1 + \frac{S(t_2)}{S(t_1)} - 1 \right)   \approx \frac{S(t_2)-S(t_1)}{S(t_1)} = R(t_2)
\end{equation}

Using the approximation in equation (\ref{eq:taylorapprox}), we can see that the returns $R(t)$ approximately follow normal distribution using the assumption of GBM for stock prices.

\subsection[Asymptotic distribution of eigenvalues ]{Asymptotic distribution of eigenvalues and Test Statistic} \label{sec:twostock}

\begin{flushleft}
	Asymptotic distribution of the eigenvalues of sample covariance matrix is derived in equations (2.1 - 2.13) of \cite{anderson1963asymptotic}. The theorem and results are as follows: 
\end{flushleft}

Say $x_{\alpha}$ is a $p$-dimensional vector distributed according to $\mathcal{N}(0,\Sigma)$ and $A = \sum_{1}^{n} x_{\alpha}x_{\alpha}^{'}$. From multivariate central limit theorem, $(1/n^{1/2})(A - n\Sigma)$ is asymptotically normally distributed. Let $(d_1,\dots,d_p)$ be the eigenvalues of $A$, where $d_1 \ge d_2 \ge \dots \ge d_p$,  and $(\delta_1,\dots,\delta_p)$ be the eigenvalues of $\Sigma$, where $\delta_1 \ge \delta_2 \dots \ge \delta_p$.
\\

First, we give the results for the simple case where all the characteristic roots of $\Sigma$ are equal, that is, $\delta_1 = \dots = \delta_p = \lambda$, say. Define  
$H := \sqrt{t}(D_t - \lambda I)$, where $D_t$ is a $p$ x $p$ diagonal matrix  with $(d_1,\dots,d_p)$ as diagonal elements, and $I$ is the $p$ x $p$ identity matrix. Clearly, $H$ is also a $p$ x $p$ diagonal matrix with elements say, $h_1, h_2, h_3, \dots, h_p$. Asymptotic distribution of $H$ is given by

\begin{align}
    f(h_1, h_2, h_3, \dots, h_p; \lambda, p) = \frac{\mathbf{K}(p)}{\lambda^{\frac{p(p+1)}{2}}} e^{-\frac{\sum_{1}^{p}h_i^2}{4(\lambda)^2}}\prod_{i<j}^{} (h_i - h_j) \qquad \label{eq:eigendisbtn}
\end{align}

where, $$\frac{1}{\mathbf{K}(p)} = 2^{\dfrac{p(p+3)}{4}}\prod_{i=1}^{p}\Gamma[\frac{1}{2}(p+1-i)]$$

$$$$

If the largest eigenvalue, $\lambda$, has a multiplicity of $q$ instead of $p$, where $q \le p$, then test statistic is slightly modified as $H = \sqrt{t}(D_t - \lambda I)$, where $D_t$ is the $q$ x $q$ sample eigenvalue diagonal matrix and $I$ is the $q$ x $q$ identity matrix. The distribution of $H$ is same as the one given above in equation (\ref{eq:eigendisbtn}) with $p$ replaced by $q$. For example, when the maximum eigenmultiplicity is just 1 (the case when all the eigenvalues are unique), our equation (\ref{eq:eigendisbtn}) simplifies to 

\begin{align}
    f (h_i; \lambda, 1) = \frac{1}{2 \sqrt \pi \lambda} e^{\frac{-h_i^2}{4(\lambda)^2}} \label{eq:simplfieddistbn}
\end{align}

\subsubsection{Simplification for two-stock portfolio} 
We can simplify the above distribution in equation (\ref{eq:eigendisbtn}) for the simple case of two-stock portfolio, i.e. $p = 2$. Let $X_\alpha = (X_t,Y_t)$, where $X_t, Y_t$ are standardized returns. Also, let $A_t = \sum_{1}^{t}X_\alpha X_{\alpha}^{'}$. 
The correlation matrix, $A_T$, at time $T$ can be represented as 

$$\begin{pmatrix}1&\rho_T\\\rho_T&1\end{pmatrix}$$ whose eigenvalues are $1+\rho_T$ and $1-\rho_T$ and the corresponding eigenvectors are $\dfrac{1}{\sqrt{2}}(1,1)$ and $\dfrac{1}{\sqrt{2}}(1,-1)$. Let $D_T = 1+\rho_T$ be its largest eigenvalue.\\

The sample correlation matrix, $A_t$, at time $t$ is $$\begin{pmatrix}1 &\rho_t\\\rho_t&1 \end{pmatrix}$$ \\

Also, let $D_t$ be the largest eigenvalue of $A_t$\\

The test statistic is defined as follows: $$h_t = \sqrt{t}(D_t - D_T)  = \sqrt{t}((1+\rho_t)-(1+\rho_T))$$
 where, $D_t , D_T$ are the maximum eigenvalues at times $t$ and $T$ respectively. Note that our test statistic is qualitatively very similar to the one defined in \cite{wied2012testing} as  $D \times \max_{2\le j \le T} \frac{j}{\sqrt{T}} |\rho_j - \rho_T|$, where $D$ is scalar constant.
 \\

From equation (\ref{eq:simplfieddistbn}), we can see that the asymptotic distribution of $h_t$ is $$h_t = \sqrt{t}((1+\rho_t)-(1+\rho_T)) \xrightarrow{d} \frac{1}{2\sqrt{\pi}}\dfrac{1}{(1+\rho_T)}e^{-\frac{h_t^2}{4(1+\rho_T)^2}} $$

\subsubsection{Simplification for three-stock portfolio}
In this subsection,  we derive an expression for the  asymptotic distribution of the test statistic for a three-stock portfolio, i.e. $p = 3$. Let $X_\alpha = (X_t,Y_t,Z_t)$, where $X_t, Y_t, Z_t$ are standardized returns. Also, let $A_t = \sum_{1}^{t}X_\alpha X_{\alpha}^{'}$. 
The correlation matrix, $A_T$, at time $T$ can be represented as 

$$\begin{pmatrix}1&\rho_{1,T}&\rho_{2,T}\\ \rho_{1,T}& 1& \rho_{3,T}\\ \rho_{2,T}&\rho_{3,T}&1\end{pmatrix}$$ 

The characteristic equation for the above matrix is \begin{align}
    (1-\lambda)^3 - (1-\lambda)(\rho_{1,T}^2+\rho_{2,T}^2+\rho_{3,T}^2) + 2\rho_{1,T}\rho_{2,T}\rho_{3,T} = 0 \label{eq:3stockcorrelation}
\end{align} 

Solving for $\lambda$ in equation (\ref{eq:3stockcorrelation}), gives the eigenvalues of the correlation matrix $A_T$.

Let $\lambda^* = 1-\lambda $, then equation (\ref{eq:3stockcorrelation}) becomes

\begin{align}({\lambda^*})^3 + p\lambda^* = q  \label{eq:cubicequation} \end{align}

where, $p = -(\rho_{1,T}^2+\rho_{2,T}^2+\rho_{3,T}^2) $ , $q = -2\rho_{1,T}\rho_{2,T}\rho_{3,T} $ 

Further, let $Q = \frac{p}{3}$, $R=\frac{q}{2}$ and $D = Q^3 + R^2$.

From AM-GM inequality, we can see that $D \le 0$. If $D \le0 $, \cite{cubicformula}'s equations (57-73) give the real roots to equation (\ref{eq:cubicequation})  as 

$$
\left\{
\begin{array}{ll}
D_{1,T} &= 2\sqrt{-Q}\cos(\frac{\theta}{3}) \\ 
D_{2,T} &= 2\sqrt{-Q}\cos(\frac{\theta+2\pi}{3}) \\ 
D_{3,T} &= 2\sqrt{-Q}\cos(\frac{\theta+4\pi}{3})
\end{array} 
\right.
$$

where $\theta = \cos^{-1}\left(\frac{R}{\sqrt{-Q^3}}\right)$\\
Similarly, we can find the roots $D_{1,t},D_{2,t},D_{3,t}$ corresponding to the correlation matrix, $A_t$, at time $t$. Let $D_{1,T}$ and $D_{1,t}$ be the smallest roots corresponding to $A_T$ and $A_t$ respectively. As this is the case of multiplicity being 1, from equation (\ref{eq:simplfieddistbn}), we can see that the asymptotic distribution of $h_t(= \sqrt{t}(D_{1,T}-D_{1,t}))$ is $$h_t \xrightarrow{d} \frac{1}{2\sqrt \pi}\dfrac{1}{\left(1-2\sqrt{-Q}\cos(\frac{\theta}{3})\right)}e^{-\frac{h_t^2}{4{ \left(1-2\sqrt{-Q}\cos(\frac{\theta}{3})\right) }^2}} $$

\section{Testing on stock returns}\label{Testing on stock returns}

\begin{figure}
	\includegraphics[width=\linewidth]{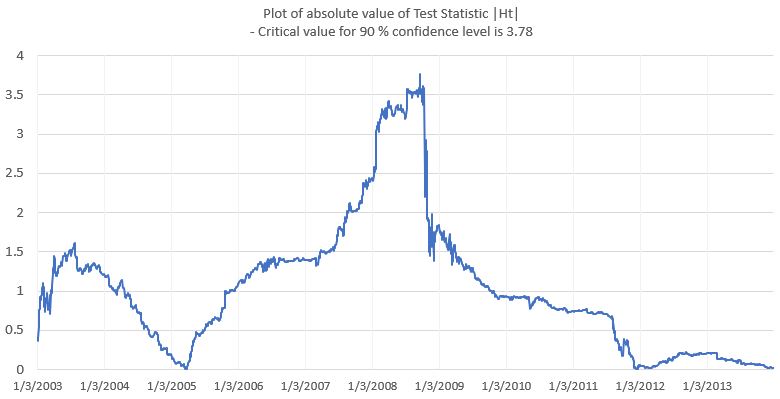}
	\caption{Absolute value of test statistic for correlation between SNP500 \& DAX indices}
	\label{fig:2stock}
\end{figure}

First, we demonstrate our test on the same stocks (SNP500 and DAX indices) considered in \cite{wied2012testing};  the highest value of the test statistic $|h_t|$ is 3.79; as seen from Figure~\ref{fig:2stock}, \textit{this coincides with the collapse of Lehman Brothers around 18 September, 2008,} and is slightly greater than the critical value for 90\% confidence level. 

\begin{figure}[H]
	\centering
	\begin{subfigure}[b]{0.9\linewidth}
		\includegraphics[width=\linewidth]{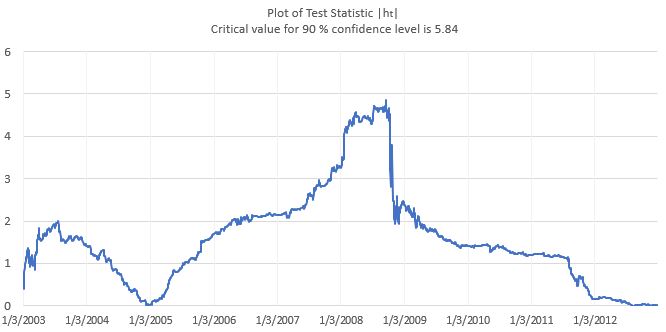}
		\caption{SNP, DOWJones, DAX}
	\end{subfigure}
	\begin{subfigure}[b]{0.9\linewidth}
		\includegraphics[width=\linewidth]{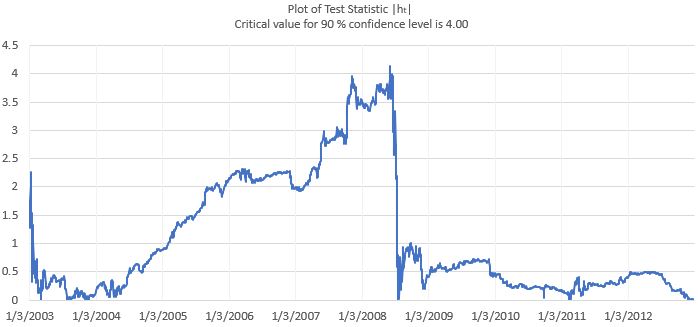}
		\caption{SNP, DAX, Nikkei.}
	\end{subfigure}
	\caption{The figure (a), as evident from critical values, does not show considerable breakdown in correlation as against the case when all the stocks are from different countries in figure (b)}
	\label{fig:coffee}
\end{figure}

Next, we demonstrate our test on a three-stock portfolio for two cases. In case (i), we have two American Indices (SNP 500 and DOWJones) and  a German Index (DAX). In case (ii), we have indices from US, Germany and Japan (SNP, DAX and Nikkei respectively). \textit{As expected, it can be seen from Figure~\ref{fig:coffee}, the correlation structure has been disturbed more in case(ii), where all the indices are from different countries,  as against case(i) where 2 indices are from the same country (US).}

\section{Local power} \label{Local power}

First, we derive the distribution of the test statistic $h_t =  \sqrt{t}((1+\rho_t)-(1+\rho_T))$ for the alternative hypothesis under consideration for cases (1-4) mentioned below, where the correlation changes once at time $t_1$. For case(5), the correlation changes twice, once at time $t_1$ and once at $t_1 + t_2$.

\begin{enumerate}
	\item $\rho_i = 0.5$, $i \le \frac{T}{2}$ and $\rho_i = 0.7$, $i > \frac{T}{2}$
	\item $\rho_i = 0.5$, $i \le \frac{T}{4}$ and $\rho_i = 0.7$, $i > \frac{T}{4}$
	\item $\rho_i = 0.5$, $i \le \frac{T}{2}$ and $\rho_i = -0.5$, $i > \frac{T}{2}$
	\item $\rho_i = 0.5$, $i \le \frac{T}{4}$ and $\rho_i = -0.5$, $i > \frac{T}{4}$
	\item $\rho_i = 0.5$, $i \le \frac{T}{4}$ and $\rho_i = 0.7$, $\frac{T}{2} <i < \frac{3T}{4}$ and $\rho_i = 0.5, \, i \ge \frac{3T}{4}$
	
\end{enumerate}

First we derive the expressions for cases (1-4) where correlation changes only once. Let $t_1$ be the time until which the correlation remains $\rho_1$ and from time $t_1$ to $t_1+t_2$ the correlation remains $\rho_2$.

So, the correlation matrices for duration $t_1$ and $t_2$ can be written as
$$ \frac{\sum_{t=1}^{t_1} x_\alpha x_\alpha^{'}}{t_1} =   E \begin{pmatrix}d_{1_{t_1}}&0\\0&d_{2_{t_1}}\end{pmatrix} E^{'} $$

$$ \frac{\sum_{t=t_1}^{t_1+t_2} x_\alpha x_\alpha^{'}}{t_2} =   E \begin{pmatrix}d_{1_{t_2}}&0\\0&d_{2_{t_2}}\end{pmatrix} E^{'} $$

	Therefore, the correlation matrix at time $t_1 + t_2$ can be written as 

$$ \frac{\sum_{t=0}^{t_1+t_2} x_\alpha x_\alpha^{'}}{t_1+t_2} =   E \begin{pmatrix}\frac{t_1(d_{1_{t_1}})+t_2(d_{1_{t_2}})}{t_1+t_2}&0\\0&\frac{t_1(d_{1_{t_1}})+t_2(d_{1_{t_2}})}{t_1+t_2}\end{pmatrix} E^{'} $$
 where $EE^' = I$

	From equation (\ref{eq:eigendisbtn}), we can get the distribution of $d_{1_{t_1}},\, d_{1_{t_2}},$ and, hence, the distribution of $d_{1_{t_1}}+d_{1_{t_2}} $.

	When the correlation changes only once at $t_1$, the form of our test statistic at $t_1$ is $$h_{t={t_1}} = \sqrt{t_1}\left(d_{1_{t_1}} - \frac{t_1(d_{1_{t_1}})+t_2(d_{1_{t_2}})}{t_1+t_2}\right) = \frac{\sqrt{t_1}t_2}{t_1+t_2}(d_{1_{t_1}}-d_{1_{t_2}})$$

Substituting the distributions of $d_{1_{t_1}}$ and $d_{1_{t_2}}$, we have,

$$  \frac{\sqrt{t_1}t_2}{t_1+t_2} \left((\rho_1 - \rho_2) + \left( \sqrt{\left( \frac{(1+\rho_1)^2}{0.5t_1}\right)  +  \left( \frac{(1+\rho_2)^2}{0.5t_2}\right) } \right) \mathbb{N}_s \right)$$

Where, $\mathbb{N}_s$ is standard normal distribution

E.g. For case (1) and $T = 100$, we have $t_1 , t_2 = 100$,  $\rho_1 = 0.5$ and $\rho_2 = 0.7$ \\

We can derive a similar expression for the case (5) where the correlation changes at two times, $t_1$ and $t_1 + t_2$, in the total duration ($t_1+t_2+t_3$) under consideration.

The test statistic at $t_1$ is

$$h_{t={t_1}} = \sqrt{t_1}\left(d_{1_{t_1}} - \frac{t_1(d_{1_{t_1}})+t_2(d_{1_{t_2}})+t_3(d_{1_{t_3}})}{t_1+t_2+t_3}\right)$$

\begin{flushleft}
	Substituting the distributions of $d_{1_{t_1}}$, $d_{1_{t_2}}$, and, $d_{1_{t_3}}$,  we have the distribution of $h_{t={t_1}}$ as 
\end{flushleft}

\begin{dmath*}
	\frac{\sqrt{t_1}}{t_1+t_2+t_3} \left( t_2\left((\rho_1 - \rho_2) + \left( \sqrt{\left( \frac{(1+\rho_1)^2}{0.5t_1}\right) +  \left( \frac{(1+\rho_2)^2}{0.5t_2}\right) } \right) \mathbb{N}_s \right)  + t_3\left((\rho_1 - \rho_3) + \left( \sqrt{\left( \frac{(1+\rho_1)^2}{0.5t_1}\right)  +  \left( \frac{(1+\rho_3)^2}{0.5t_3}\right) } \right) \mathbb{N}_s \right)  \right) 
\end{dmath*}

We checked the power of our test for cases (1-5), in which variances constantly remain 1 and correlations change, and compared our results against those of \cite{wied2012testing} and \cite{aue2009break}.

\begin{table}[H]
	\centering
	\caption{Empirical size-adjusted rejection frequencies when correlations change for cases (1-5) with varying time horizon T}
	\begin{tabular}{|rrrrrrr|}
		\hline
		T     &       & 1     & 2     & 3     & 4     & 5 \\
		\hline
		\hline
		\multicolumn{6}{c}{(a) Our test}         \\
		\hline
		200   &       & 0.01 & 0.04 & 0.70   & 0.70   & 0.04 \\
		500   &       & 0.04 & 0.07  & 0.99 & \textbf{0.99} & 0.06 \\
		1000  &       & 0.08 & 0.14  & \textbf{1}     & \textbf{1}     & 0.08 \\
		2000  &       & 0.21  & 0.29  & \textbf{1}     & \textbf{1}     & 0.15 \\
	\hline
		\multicolumn{6}{c}{(b) Test of [Wied 2012]}  \\
		\hline
		200   &       & 0.309 & 0.255 & 0.953 & 0.88  & 0.101 \\
		500   &       & 0.255 & 0.488 & 0.996 & 0.989 & 0.207 \\
		1000  &       & 0.83  & 0.733 & 0.998 & 0.998 & 0.422 \\
		2000  &       & 0.967 & 0.928 & 1     & 0.999 & 0.75 \\
	\hline
		\multicolumn{6}{c}{(c) Test of [Aue 2009]}  \\
		\hline
		200   &       & 0.24 & 0.16 & 0.97 & 0.84 & 0.08 \\
		500   &       & 0.58 & 0.40   & 1     & 0.99 & 0.16 \\
		1000  &       & 0.85 & 0.69  & 1     & 1     & 0.28 \\
		2000  &       & 0.98  & 0.93 & 1     & 1     & 0.61 \\
		\hline
	\end{tabular}%
	\label{tab:addlabel}%
\end{table}%

\begin{figure}[H]
	\centering
	\begin{subfigure}[b]{0.9\linewidth}
		\includegraphics[width=\linewidth]{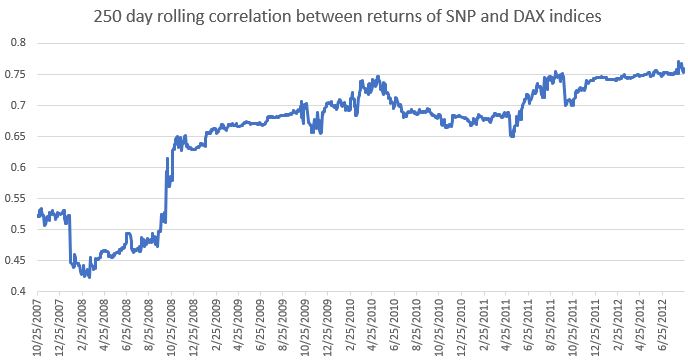}
		\caption{SNP and DAX}
	\end{subfigure}
	\begin{subfigure}[b]{0.9\linewidth}
		\includegraphics[width=\linewidth]{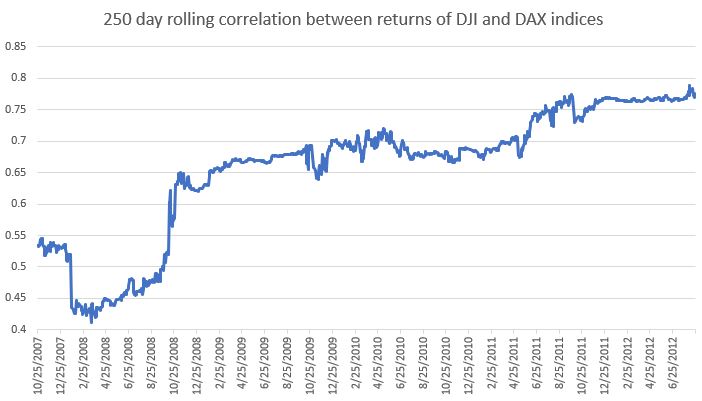}
		\caption{DOW Jones and DAX }
	\end{subfigure}
	\caption{During the times of financial crisis, it is common to observe a change in correlation of about (0.3 - 0.35) in a two stock portfolio}
	\label{fig:coffee1}
\end{figure}

From Table~\ref{tab:addlabel}, it is seen that, in general, the power of our test is lower compared to those of \cite{wied2012testing} and \cite{aue2009break}. However, in particular, for cases 3 and 4, the power of our test is slightly higher - indicating that our test can detect large changes in the correlation structure more effectively. Further, it should be noted that a correlation change of about (0.3 - 0.35) is common during the times of financial crisis as indicated in Figure~\ref{fig:coffee1}. So, while the power of our test is comparatively lower, our approach is simpler to explain, understand \& implement, and still can be used to test the breakdown in correlation structure for real world scenarios. 

\section{Conclusions}\label{conclusions}
In this paper, we have proposed a new fluctuation test for constant correlation matrix under a multivariate setting in which the change points need not be specified apriori. Our approach is more simplified because it allows us to work with more standard operations like eigen-decomposition and normal distributions against the pairwise comparison and Brownian bridges of \cite{wied2012testing} and \cite{wied2017nonparametric}. Since we are dealing with only the largest eigenvalue, the power of our test is lower compared to the pairwise comparison of the entire correlation matrix of \cite{wied2012testing} and \cite{wied2017nonparametric}. Nevertheless, our test is  simpler in terms of understanding and practical application, and is effectively able to detect changes in correlation matrix in real world scenarios, as indicated in Section \ref{Testing on stock returns}.  Moreover, our method can be generalized to detect any changes in covariance matrix structure, as a complementary technique to \cite{aue2009break}. One drawback of our test, which is also shared by \cite{wied2012testing} and \cite{wied2017nonparametric}, is the assumption of finite fourth moments and constant expectations and variances. Another drawback, which is shared by most of correlation based tests, is the low power when there are multiple change points in the duration under consideration, as illustrated in \cite{cabrieto2018testing}. Hence, it may be worthwhile to consider techniques like prefiltering and/or other transformations to overcome these drawbacks. 

\bibliographystyle{apalike}

\end{document}